\documentclass[12pt]{iopart}

\usepackage{iopams}

\usepackage{graphicx}

\newcommand{\dd}{{\textrm d}}
\newcommand{\GeV}{{\textrm{GeV}}}
\newcommand{\MeV}{{\textrm{MeV}}}
\newcommand{\cm}{{\textrm{cm}}}
\newcommand{\fm}{{\textrm{fm}}}
\renewcommand\theenumi{\roman{enumi}}

\begin{document}

\title{Vacuum Energy Density in the Quantum Yang\,--\,Mills Theory }

\author{G.G. Barnaf\"{o}ldi$^{1,2}$ and V. Gogokhia$^1$}

\vspace{3mm}

\address{$^1$HAS, CRIP, RMKI, Department of Theoretical Physics,
Budapest 114, P.O.B. 49, H-1525, Hungary}

\address{$^2$Center for Nuclear Research, Department of Physics,
Kent State University, Kent, Ohio 44242, USA}

\eads{\mailto{bgergely@rmki.kfki.hu}, \mailto{gogohia@rmki.kfki.hu}}

\date{\today}

\begin{abstract}
Using the effective potential approach for composite operators, we
have formulated a general method of calculation of the truly
non-perturbative Yang-Mills vacuum energy density (this is, by
definition, the bag constant apart from the sign). It is the main
dynamical characteristic of the QCD ground state. Our method
allows one to make it free of the perturbative contributions
('contaminations'), by construction. We also perform an actual
numerical calculation of the bag constant for the confining effective
charge. Its choice uniquely defines the bag constant, which becomes
free of all the types of the perturbative contributions now, as well
as possessing many other desirable properties as colorless, gauge
independence, etc. Using further the trace anomaly relation, we
develop a general formalism which makes it possible to relate the
bag constant to the gluon condensate defined at the same $\beta$
function (or, equivalently, effective charge) which has been chosen for the calculation
of the bag constant itself. Our numerical result for it shows a good agreement with other
phenomenological estimates of the gluon condensate. We have argued that the calculated bag constant
may contribute to the dark energy density. Its contribution is by 10 orders of magnitude
better than the estimate from the Higgs field's contribution. We also propose to consider the bag energy as
a possible amount of energy which can be released from the QCD ground state by a single cycle. The QCD ground state is shown to be an infinite and hence a permanent reservoir of energy.
\end{abstract}

\pacs{ 11.15.Tk, 12.38.Lg, 12.38.Aw}


\maketitle
\section{Introduction}
\label{sec:intro}

In order to calculate physical observables from first principles in
Quantum Chromodynamics (QCD) \cite{1} it is not enough to know its
Lagrangian. It is also necessary and important to know the true
structure of its ground state. It is the response of the QCD vacuum
which substantially modifies all the QCD Green's functions from
their free counterparts. These full ("dressed") Green's functions
are needed for the above-mentioned calculations. The vacuum of QCD
is a very complicated confining medium and its dynamical and
topological complexity means that its structure can be organized at
various levels: classical and quantum \cite{1,2,3,4,5,6} (and
references therein). It is mainly non-perturbative (NP) by origin,
character and magnitude, since the corresponding coupling constant
is large. However, the virtual gluon field configurations and
excitations of the perturbative (PT) origin, character and
magnitude, due to asymptotic freedom (AF) \cite{1}, are also
present there.

One of the main dynamical characteristics of the QCD ground state
is the bag constant. Its name comes from the famous bag models for hadrons
\cite{7,8}, but its present understanding (and thus modern
definition) not connecting to hadron properties. It is defined as
the difference between the PT and the NP vacuum energy densities
(VEDs) \cite{9,10,11,12}. So, we can symbolically put
$B = VED^{PT} - VED$, where $VED$ is the NP but 'contaminated' by
the PT contributions (i.e., this is a full $VED$ like the full
gluon propagator, see below). At the same time, we can continue as
follows: $B = VED^{PT} - VED =
VED^{PT} - [VED - VED^{PT} + VED^{PT}] = VED^{PT}
- [VED^{TNP} + VED^{PT}] = - VED^{TNP} > 0$, since the VED is always
negative. The bag constant is nothing but the truly NP (TNP) VED,
apart from the sign, by definition, and thus is free of the PT
contributions ('contaminations'). The symbolic subtraction
presented here includes the subtraction at the fundamental gluon
level, and two others at the hadronic level, i.e., when the gluon
degrees of freedom should be integrated out (see section
\ref{sec:tnpved} below). In order to consider it also as a physical
characteristic of the QCD ground state, the bag constant correctly
calculated should satisfy some other necessary requirements such
as colorlessness, finiteness, gauge-independence, no imaginary
part (stable vacuum), etc.

The main purpose of this paper is to formulate a formalism how to
calculate correctly the quantum part of the bag constant, using
the effective potential approach for composite operators
\cite{13,14,15}. In particular, to show how the above-mentioned
subtractions are to be analytically made. On account of the
confining effective charge, the bag constant has been numerically
evaluated, satisfying all the necessary requirements mentioned
above. Using further the trace anomaly relation \cite{16,17,18,19},
we also develop a general formalism which makes it possible to
relate the bag constant to another important NP characteristic of
the QCD ground state - the gluon condensate \cite{11}. Here, we do
not use the weak coupling solution for the corresponding $\beta$
function. Finally we present our numerical result for the bag
constant, which is in a good agreement with other phenomenological
estimates of the gluon condensate ~\cite{11,20}.

\section{The VED}
\label{sec:ved}

\begin{figure}[h]
\begin{center}
\includegraphics{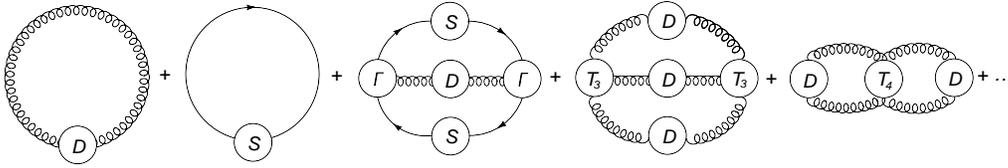}
 \caption{The skeleton loop expansion for the
effective potential. The wavy lines describe the full gluon
propagators $D$. The solid lines describe the full quark
propagators $S$. $\Gamma$ is the full quark-gluon vertex, while
$T_3$ and $T_4$ are the full three- and four-gluon vertices,
respectively.}
\label{fig:1}
\end{center}
\end{figure}

\begin{figure}[h]
\begin{center}
\includegraphics{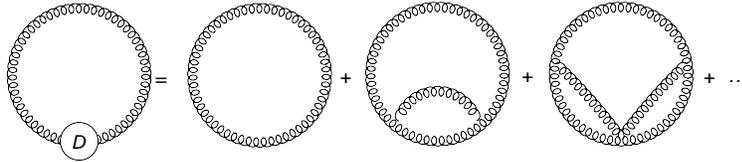}
\caption{Infinite series for the gluon part of the VED (taking
the first skeleton diagram in Fig.~\ref{fig:1}). }
\label{fig:2}
\end{center}
\end{figure}

The quantum part of the VED is determined by the effective
potential approach for composite operators \cite{13,14,15}. In the
absence of external sources the effective potential is nothing but
the VED. It is given in the form of the skeleton loop expansion
containing all the types of the QCD full propagators and vertices,
see Fig. \ref{fig:1}. So each vacuum skeleton loop itself is a sum
of an infinite number of the corresponding PT vacuum loops (i.e.,
containing the point-like vertices and free propagators, see Fig.
\ref{fig:2}, where one term in each lower order is shown, for
simplicity). The number of the vacuum skeleton loops goes with
the power of the Planck constant, $\hbar$.

Here we are going to formulate a general method of numerical
calculation of the quantum part of the TNP Yang-Mills (YM)
VED in the covariant gauge QCD. The gluon part of the VED to
leading order (the so-called log-loop level $\sim \hbar$, the
first skeleton loop diagram in Fig.~\ref{fig:1}, and which PT
expansion is shown explicitly in Fig.~\ref{fig:2}) is analytically
given by the effective potential for composite operators as
follows \cite{13}:

\begin{equation}\label{eq:1}
V(D) =  { i \over 2} \int {\dd ^4q \over {(2\pi)^4}}
 \Tr \left\{ \ln (D_0^{-1}D) - (D_0^{-1}D) + 1 \right\},
\end{equation}
where $D(q)$ is the full gluon propagator and $D_0(q)$ is its free
counterpart (see below). The traces over space-time and color
group indices are assumed. Evidently, the effective potential is
normalized to $V(D_0) = 0$, i.e., the free PT vacuum is normalized
to zero, as usual. Next-to-leading and higher contributions (two
and more vacuum skeleton loops) are suppressed at least by one
order of magnitude in powers of $\hbar$. They generate very small
numerical corrections to the log-loop terms, and thus are not
important for the numerical calculation of the bag constant to
leading order.

The two-point Green's function, describing the full gluon
propagator, is
\begin{equation}\label{eq:2}
D_{\mu\nu}(q) = - i \left\{ T_{\mu\nu}(q)d(-q^2; \xi) + \xi
L_{\mu\nu}(q) \right\} {1 \over q^2 },
\end{equation}
where $d(-q^2; \xi)$ is the gluon invariant function
(dimensionless), the so-called Lorentz structure (sometimes, we
will call it as the full gluon form factor or, equivalently,
the effective charge ("running"), see below), while $\xi$ is the
gauge-fixing parameter and

\begin{equation}\label{eq:3}
T_{\mu\nu}(q) = g_{\mu\nu} - {q_\mu q_\nu \over q^2} = g_{\mu\nu }
- L_{\mu\nu}(q).
\end{equation}
Its free PT counterpart $D_0 \equiv D^0_{\mu\nu}(q)$ is obtained by
putting the full gluon form factor $d(-q^2; \xi)$ in Eq.~(\ref{eq:2})
simply to one, i.e.,

\begin{equation}\label{eq:4}
D^0_{\mu\nu}(q) = - i \left\{ T_{\mu\nu}(q) + \xi L_{\mu\nu}(q)
\right\} {1 \over q^2}.
\end{equation}

In order to evaluate the effective potential (\ref{eq:1}), on
account of Eq.~(\ref{eq:2}), we use the well-known expression

\begin{equation}\label{eq:5}
\Tr \ln (D_0^{-1}D) = 8 \times 4 \ln \det (D_0^{-1}D) = 32 \ln [ (3/
4 )d(-q^2; \xi) + (1 / 4 ) ],
\end{equation}
which becomes zero indeed when setting $d(-q^2; \xi)=1$.

Going over to four-dimensional Euclidean space in Eq.~(\ref{eq:1}),
one obtains ($\epsilon_g = V(D)$)

\begin{equation}\label{eq:6}
\epsilon_g = - 16 \int {\dd^4q \over (2\pi)^4} \left[ \ln [1 + 3
d(q^2; \xi)] - {3 \over 4}d(q^2; \xi) + a \right],
\end{equation}
where constant $a = (3/4) - 2 \ln 2 = - 0.6363$ and the
integration from zero to infinity over $q^2$ is assumed. The VED
$\epsilon_g$ derived in Eq.~(\ref{eq:6}) is already a colorless
quantity, since it has been summed over color indices. Also it does
not depend explicitly on the unphysical (longitudinal) part of the
full gluon propagator due to the product $(D_0^{-1}D)$, which, in
its turn, comes from the above-mentioned normalization to zero.
Thus it is worth emphasizing that the transversal ("physical")
degrees of freedom only of gauge bosons contribute to this
equation. Note, in the effective potential approach to leading order there is no
need for ghost degrees of freedom from the very beginning in order
to cancel the longitudinal ("unphysical") component of the full
gluon propagator. This role is played by the normalization
condition (that is why the ghost skeleton loops are not
shown in Fig.~\ref{fig:1}). Furthermore, overall numerical factor $1/2$ has been introduced
into Eq. (1) in order to make the gluon degrees of freedom to be
equal $32/2=16 = 8 \times 2$, where $8$ color of gluons times $2$
helicity (transversal) degrees of freedom, see Eqs.~(\ref{eq:5})
and~(\ref{eq:6}).

In the connection with the above-mentioned normalization condition a few remarks are in order. It does not work for the higher order vacuum loops. As explained in Ref. \cite{13}, for consistency with them in the
PT QCD Green's functions, for example in the Hartree-Fock approximation, the Landau gauge should be used.
In Ref. \cite{Castorina85} the effective potential has been used to the two-loop order for the investigation of QCD chiral-symmetry breaking just in the Landau gauge and in the Hartree-Fock approximatiion.
In the general case (i.e., beyond the PT and at any gauge), however, the cancelation of unphysical gluon modes should proceed with the help of ghosts as it is described in more detail in appendix A.

The derived expression~(\ref{eq:6}) remains rather formal, since
it suffers from the two serious problems: the coefficient of the
transversal Lorentz structure $d(q^2; \xi)$ may still depend
explicitly on $\xi$. Furthermore, it is divergent at least as the
fourth power of the ultraviolet (UV) cutoff, and therefore suffers
from different types of the PT contributions.

\section{The TNP VED}
\label{sec:tnpved}

In order to define the VED free of the above-mentioned PT
contributions ('contaminations'), let us make first the subtraction
at the fundamental gluon level, namely
\begin{equation}\label{eq:7}
d(q^2; \xi) = d(q^2; \xi) - d^{PT}(q^2; \xi)  + d^{PT}(q^2; \xi) =
d^{TNP}(q^2)  + d^{PT}(q^2; \xi),
\end{equation}
where $d^{PT}(q^2; \xi)$ correctly describes the PT structure of
the full effective charge $d(q^2; \xi)$, including its behavior in
the UV limit (AF, \cite{1}), otherwise remaining arbitrary. On the
other hand, $d^{TNP}(q^2)$ defined by the above-made subtraction,
is assumed to reproduce correctly the TNP structure of the full
effective charge, including its asymptotic in the deep infrared
(IR) limit. This underlines the strong intrinsic influence of the
IR properties of the theory on its TNP dynamics. Evidently, both
terms are valid in the whole energy/momentum range, i.e, they are
not asymptotics. Let us also emphasize the principle difference
between $d(q^2; \xi)$ and $d^{TNP}(q^2)$. The former is the NP
quantity "contaminated" by the PT contributions, while the latter
one being also NP, nevertheless, is free of them. Thus the formal
separation between the TNP effective charge $d^{TNP}(q^2)$ and
its PT counterpart $d^{PT}(q^2; \xi)$ is achieved.
For example, if the full effective charge explicitly depends on the scale
responsible for the TNP dynamics in QCD, say $\Delta^2$ - the
so-called mass gap (see section~\ref{sec:effq} below), then one
can define the subtraction as follows:
$d^{TNP}(q^2; \Delta^2) = d(q^2; \Delta^2) - d(q^2; \Delta^2=0) =
d(q^2; \Delta^2)  - d^{PT}(q^2)$, which is, obviously, equivalent
to the decomposition~(\ref{eq:7}). In this way the separation
between the TNP effective charge and its PT counterpart becomes
exact, but not unique. Let us emphasize that the dependence of
the full effective charge $d(q^2, \Delta^2)$ on $\Delta^2$ can
be only regular. Otherwise it is impossible to assign to it the
above-mentioned physical meaning, since $\Delta^2$ can be only
zero (the formal PT limit) or finite, i.e., it cannot be
infinitely large. In principle, in some special models of the
QCD vacuum, such as the Abelian Higgs model~\cite{22,23}, the
NP scale is to be identified with the mass of the dual gauge boson.

There is also another serious reason for the subtraction
in Eq.~(\ref{eq:7}). The problem is that the above-mentioned UV
asymptotic of the full effective charge may depend on the
gauge-fixing parameter $\xi$ explicitly, namely to leading order
$d(q^2, \xi) \sim_{q^2 \rightarrow \infty} \big( \ln (q^2 /
\Lambda^2_{QCD}) \big) ^{c_0/b_0}$, where the exponent
$(c_0/b_0) < 0$ explicitly depends on the gauge-fixing parameter
$\xi$ via the coefficient $c_0$ based on Ref.~\cite{1},
and $\Lambda^2_{QCD}$ is the QCD asymptotic scale parameter. In
this connection let us note that AF being a physical phenomenon
does not depend on the gauge choice (it takes place at any gauge),
while the UV asymptotic of the corresponding Green's function may
be still gauge-dependent. This is just explicitly shown above.
Evidently, in the decomposition~(\ref{eq:7}) just the PT part of
the full effective charge will be responsible for this explicit
dependence on the gauge choice. Subtracting it, we will be
guaranteed that the remaining part will not depend explicitly on
the gauge-fixing parameter (that is why the dependence on $\xi$
is not explicitly shown in $d^{TNP}(q^2)$). Let us note that if
there is no exact criterion how to distinguish between the TNP
and the PT parts in the full effective charge in Eq.~(\ref{eq:7})
as described above, then it is possible from the full effective
charge to subtract its UV asymptotic only. However, in this case
the separation between the TNP and the PT parts will be neither
exact nor unique. For how to make this separation exact and
unique at the same time see section~\ref{sec:effq}.

Substituting the decomposition~(\ref{eq:7}) into Eq.~(\ref{eq:6})
and doing some simple rearrangements, one obtains

\begin{equation}\label{eq:8}
\epsilon_g = - {1 \over \pi^2} \int \dd q^2 \ q^2 \left[ \ln [1 +
3 d^{TNP}(q^2)] - {3 \over 4}d^{TNP}(q^2) \right] + \epsilon_{PT},
\end{equation}
where the trivial integration over the angular variables in Eq. (6) has been already done.
Here $\epsilon_{PT}$ is

\begin{equation}\label{eq:9}
\epsilon_{PT} = - {1 \over \pi^2}  \int \dd q^2 \ q^2 \left[ \ln
[1 + {3d^{PT}(q^2; \xi) \over 1 + 3 d^{TNP}(q^2)}] - {3 \over
4}d^{PT}(q^2; \xi) + a \right].
\end{equation}
It contains the contribution which is mainly determined by the PT
part of the full effective charge, $d^{PT}(q^2, \xi)$. The
constant $a$ should be also included, since it comes from the
normalization of the free PT vacuum to zero.

However, this is not the whole story yet. The first term in Eq. (8), depending only
on the TNP effective charge, nevertheless, assumes the
integration over the PT region up to infinity. It also represents the type of the PT contribution,
which should be subtracted as well. If we separate the NP region from the PT one, by introducing the
so-called effective scale $q_{eff}^2$ explicitly, then we get

\begin{equation}\label{eq:10}
\epsilon_g = - {1 \over \pi^2} \int_0^{q^2_{eff}} \dd q^2 \ q^2
\left[ \ln [1 + 3 d^{TNP}(q^2)] - {3 \over 4}d^{TNP}(q^2) \right] +
\epsilon_{PT} + \epsilon'_{PT}\ , \ \ \ \
\end{equation}
where
\begin{equation}\label{eq:11}
\epsilon'_{PT} = - {1 \over \pi^2} \int_{q^2_{eff}}^{\infty} \dd
q^2 \ q^2 \left[ \ln [1 + 3 d^{TNP}(q^2)] - {3 \over 4}d^{TNP}(q^2)
\right].
\end{equation}
This integral represents the contribution to the VED which is
determined by the TNP part of the full gluon propagator but
integrated out over the PT region. Along with $\epsilon_{PT}$
given in Eq.~(\ref{eq:9}) it also represents a type of the PT
contribution into the gluon part of the VED~(\ref{eq:8}), as
mentioned above. This means that the two remaining terms in
Eq.~(\ref{eq:10}) should be subtracted by introducing the TNP
YM VED $\epsilon_{YM}$ as follows:

\begin{equation}\label{eq:12}
\epsilon_{YM} = \epsilon_g - \epsilon_{PT} - \epsilon'_{PT},
\end{equation}
where the explicit expression for $\epsilon_{YM}$ is given by the
integral in Eq.~(\ref{eq:10}).

Concluding, let us emphasize that both subtracted terms
$\epsilon_{PT}$ and $\epsilon'_{PT}$, strictly speaking, are not
the purely PT, since along with the nontrivial PT effective charge
$d^{PT}(q^2)$ they contain the TNP effective charge
$d^{TNP}(q^2)$ as well. So to call them the PT contributions is a
convention. More precisely it is better to say that these terms
are "contaminated" by the PT contributions. The above-mentioned
necessary subtractions can be made in a more sophisticated way by
introducing explicitly the ghost degrees of freedom (see appendix A).

\section{The bag constant}
\label{sec:bag}

The bag constant (the so-called bag pressure) is defined as the
difference between the PT and the NP VEDs \cite{9,10,11,12}. So
in our notations for the YM fields, and as it follows from the
definition by Eq.~(\ref{eq:12}), it is nothing but the TNP YM VED
apart from the sign, i.e.,
\begin{eqnarray}\label{eq:13}
B_{YM} = - \epsilon_{YM} &=& \epsilon_{PT} + \epsilon'_{PT} -
\epsilon_g = \nonumber\\
&=& {1 \over \pi^2} \int_0^{q^2_{eff}} \dd q^2 \ q^2 \left[ \ln [1
+ 3 \alpha_s^{TNP}(q^2)] - {3 \over 4} \alpha_s^{TNP}(q^2) \right],
\end{eqnarray}
where from now on we introduce the notation
\begin{equation}\label{eq:14}
d^{TNP}(q^2) \equiv \alpha^{TNP}_s(q^2),
\end{equation}
since $d^{TNP}(q^2)$ is the TNP effective charge
$\alpha^{TNP}_s(q^2)$, as noted above. This is a general
expression for any model effective charge in order to calculate
the bag constant, or the TNP YM VED apart from the sign, from
first principles. It is our definition of the TNP YM VED and
thus of the bag constant. So it is defined as the special
function of the TNP effective charge integrated out over the NP
region (soft momentum region, $0 \leq q^2 \leq q^2_{eff}$). It is
free of the PT contributions, by construction. In this connection,
let us recall that $\epsilon_g$ is also NP, but 'contaminated'
by the PT contributions, which just to be
subtracted in order to get Eq.~(\ref{eq:13}) from Eq.~(10).

Comparing expressions~(\ref{eq:6}) and~(\ref{eq:13}), one comes to
the following {\it 'prescription'} to get Eq.~(\ref{eq:13}) directly
from Eq.~(\ref{eq:6}):

\begin{enumerate}
\item Replacing $d(q^2) \rightarrow d^{TNP}(q^2)$ or equivalently,
$\alpha_s(q^2) \rightarrow \alpha_s^{TNP}(q^2)$.

\item Omitting the constant $a$ which normalizes the free PT
vacuum to zero.

\item Introducing the effective scale $q^2_{eff}$ which separates
the NP region from the PT one in the $q^2$-momentum space.

\item Omitting the minus sign for the bag constant.
\end{enumerate}

At this stage the bag constant defined by Eq.~(\ref{eq:13}) is
definitely colorless (color-singlet) and free of the PT contributions ("contaminations").
Let us remind that it also depends on only transversal degrees
of freedom of gauge bosons (gluons). All its other properties
mentioned above (finiteness, positivity, no imaginary part, etc.)
depend on the chosen effective charge, more precisely on its
TNP counterpart. It is worth emphasizing once more that in defining
correctly the bag constant,
three types of the corresponding subtractions have been
introduced. The first one - in Eq.~(\ref{eq:7}) at the fundamental
gluon level and the two others - in Eq.~(\ref{eq:12}), when the
gluon degrees of freedom were to be integrated out.

For actual numerical calculations of the bag constant via the
expression~(\ref{eq:13}) it is always convenient to factorize its
scale dependence. For this purpose, let us introduce the
dimensionless variable and the TNP effective charge as follows:

\begin{equation}\label{eq:15}
\alpha_s^{TNP}(q^2)= \alpha_s^{TNP}(z), \ \ \ \textrm{where} \ \ \
z = {q^2 \over q^2_{eff}} \ .
\end{equation}
From the general expression for the bag constant~(\ref{eq:13})
in these terms one then gets

\begin{equation}\label{eq:16}
B_{YM} (q^2_{eff}) = q^4_{eff} \times \Omega_{YM},
\end{equation}
where we introduce the dimensionless TNP YM effective potential
$\Omega_{YM}$, for convenience. Its explicit expression is

\begin{equation}\label{eq:17}
\Omega_{YM}  = {1 \over \pi^2} \int_0^1 \dd z \ z \left[ \ln [1 +
3 \alpha_s^{TNP}(z)] - {3 \over 4} \alpha_s^{TNP}(z) \right].
\end{equation}
Let us emphasize that in order to factorize the scale dependence
in the effective potential it is necessary to choose the fixed
scale, like $q^2_{eff}$, and not the scale which can be varied,
for example like the mass gap which can go to zero in order to
recover the PT limit (see section below). Eqs.~(\ref{eq:16})
and~(\ref{eq:17}) are the main subject of our consideration in
what follows. It is worth emphasizing once more that these
expressions are general ones in order to correctly calculate the
Bag constant from first principles in any model gluon propagator. The only
problem remaining to solve is to choose such TNP effective charge
$\alpha_s^{TNP}(z)$ which, first of all should not {\it explicitly}
depend on the gauge-fixing parameter $\xi$. At the same time, the
implicit gauge dependence is not a problem. Such kind of the
dependence is unavoidable in quantum or classical gauge
theories, since the fields themselves are gauge-dependent
\cite{1,2}. For the different TNP effective charges
$\alpha_s^{TNP}(z)$ one gets different numerical results. That is why the choice for its explicit expression
(ansatz) should be physically and mathematically well justified (see below).

In this connection, let us remind that the gluon Schwinger\,--\,Dyson (SD)
equation is highly non-linear one, and it has a very complicated mathematical structure, so there is no hope for an exact solutions, the number of which is not even fixed ~\cite{1}. This means that the number of
independent solutions, obtained under specific truncation/approximation schemes and gauges, is not fixed
$a \ priori$ as well. From the very beginning they should be considered on equal footing.

\section{Confining effective charge}
\label{sec:effq}

Let us choose the TNP effective charge as follows:
\begin{equation}\label{eq:18}
\alpha_s^{TNP}(q^2) \ \ \longrightarrow \ \
\alpha_s^{INP}(q^2) = {\Delta^2 \over q^2},
\end{equation}
where the superscript "INP" stands for the intrinsically NP effective
charge (for a such replacement see remarks below). Here
$\Delta^2 \equiv \Delta^2_{JW}$ is the so-called Jaffe-Witten (JW)
mass gap, which is responsible for the large-scale structure of the
QCD vacuum, and thus for its INP dynamics \cite{24}. Let us note,
that how the mass gap appears in QCD has been explicitly shown in our
recent work in Ref.~\cite{25}.
\begin{itemize}

\item The gauge independence is obvious, i.e., it does not depend
explicitly on the gauge choice, since the mass gap is already
renormalized, and hence it is a finite quantity.

\item It satisfies the Wilson criterion of confinement -- area law
for heavy quarks \cite{26,27} or, equivalently, leads to the linear
rising potential between heavy quarks \cite{28,29} in continuous QCD,
"seen" also by lattice QCD \cite{30,31}. In this connection a few
remarks are in order. In the case of heavy quarks the response of
the vacuum can be neglected, and therefore the interaction between
them and gluons effectively becomes point-pike. Just this makes it
possible to describe confinement of heavy quarks in terms of the
linear rising potential, derived on the basis of the
expression~(\ref{eq:18}). For the light quarks the response of the
vacuum cannot be neglected. The corresponding quark-gluon vertex is
not point-like, and therefore there is no way to analyze confinement
of light quarks in terms of the linear rising potential. However,
the expression~(\ref{eq:18}) can be still used for the solution of
the SD equation for the quark propagator together
with the corresponding Slavnov-Taylor (ST) identity for the
vertex~\cite{32}. Confinement of light quarks is due to the analytical
properties of the corresponding Green's functions (unlike the electron
propagator, the quark propagator should have no imaginary part).
This is a principle difference in the description of confinement for
light and heavy quarks.

\item The functional dependence in the confining expression~(\ref{eq:18})
is, of course, the same for the YM fields and the full QCD. The
dependence on the number of flavors can appear only in the mass gap.

\item  It is exactly defined, since in the formal PT limit
($\Delta^2 = 0$) the INP effective charge~(\ref{eq:18}) vanishes, and hence the bag constant itself.

\item It is uniquely defined as well. In order to show this explicitly,
let us assume that it can be replaced by some arbitrary function as
follows:
\begin{equation}\label{eq:19}
\alpha_s^{INP}(q^2; \Delta^2) \ \ \longrightarrow \ \ {\Delta^2 \over q^2}
\times f(q^2; \Delta^2),
\end{equation}
where $f(q^2; \Delta^2)$ is the dimensionless arbitrary function,
which is regular at zero in order not to change confining
properties of the INP effective charge (18). In this case it can
be expand in Taylor series around small $q^2$, so one obtains
$f(q^2; \Delta^2) = f(0) + (q^2 / M^2) f'(q^2; \Delta^2) + ...$,
where $M^2$ is some auxiliary mass squared. Then the INP effective
charge in Eq.~(\ref{eq:19}) becomes
\begin{equation}\label{eq:20}
\alpha_s^{INP}(q^2; \Delta^2) = {\Delta^2 \over q^2} f(0) +
{\Delta^2 \over M^2} f'(q^2; \Delta^2) + ...,
\end{equation}
and substituting this into the general decomposition~(\ref{eq:7}),
one finally obtains
\begin{equation}\label{eq:21}
\alpha_s(q^2; \Delta^2) = \alpha_s^{INP}(q^2; \Delta^2) +
\alpha_s^{PT}(q^2) = {\Delta^2 \over q^2} + \alpha_s^{PT}(q^2;
\Delta^2),
\end{equation}
where not loosing generality we include the finite number $f(0)$
into the mass gap, and retaining the same notation, for simplicity.
The uniqueness is achieved at the expense of the PT effective
charge, which now becomes regularly dependent on the mass gap
(compare with the expression~(\ref{eq:9})). Evidently, the uniqueness
is due to the singular at origin structure of the INP effective charge
in Eq.~(\ref{eq:18}). In Ref.~\cite{33} it has been explicitly shown
that the TNP part of the full gluon propagator as a function of the
mass gap contains a regular at origin term as well. That is why it is
not uniquely separated from the PT gluon propagator which effective
charge is always regular at origin. We distinguish between the INP and the PT effective charges
not only by the explicit presence of the mass gap, but by the character of the
IR singularities as well \cite{33}. So only after the replacement of
Eq.~(\ref{eq:18}) the obtained expression for the bag
constant~(\ref{eq:13}) {\it becomes free of all the types of the PT
contributions ('contaminations'), indeed}.

\item In our recent work~\cite{34} we have shown that the so-called INP
gluon propagator is the purely transversal in a gauge invariant way, by
construction. It exactly converges to the gluon propagator, which
effective charge is in Eq.~(\ref{eq:18}), after the renormalization of
the mass gap is completed. For preliminary analytical investigation of
such behavior see Refs.~\cite{35,36} as well (and references therein).
Thus, we consider the expression~(\ref{eq:18}) not only as physically
and mathematically well confirmed but as uniquely justified within the
confining INP QCD~\cite{34} with its own mass gap identified with the JW mass gap for the pure
YM fields (see above).

\item There also exist direct lattice evidences that the zero momentum modes are enhanced in
the full gluon propagator (and hence in its effective charge)
\cite{Michael95,Damm98,Burgio99,Burgio98} (and references therein). A NP finite-size scaling technique was used in Ref. \cite{Lusher94} to study the evolution of the running coupling in the SU(3) YM lattice theory. At low energies it is shown to grow. The chosen analytical ansatz (18) can be considered as useful functional parametrization of these lattice results, indeed, while the scale of the enhancement is taken into account by the mass gap.

\item It is worth noting in advance that one of the attractive additional features of Eq.~(18) is
that it allows one to perform an analytical summation over the Matsubara frequencies in the generalization of the expression for the bag constant to non-zero temperatures. In this case one obtains the curve of the gluon matter pressure as a function of temperature. It and all other its derivatives (entropy and energy densities, etc.) then can be directly compared with the corresponding thermal lattice QCD curves \cite{Boyd96,Panero09}. This will make it possible for better understanding of the thermodynamical structure of the gluon matter (work in progress and preliminary numerical results are very encouraging).

\end{itemize}

In conclusion, one may consider the expression~(\ref{eq:18}) as the confining ansatz,
for simplicity. However, it is worth emphasizing that only it
satisfies all the necessary conditions discussed above. Let us also
note that for the theoretical and numerical results,
depending on the confining effective charge, see discussion in
section~\ref{sec:concl}.

\section{Analytical and numerical evaluation of the bag constant }
\label{sec:numbag}

In terms of the variable in Eq.~(\ref{eq:15}) for the INP effective
charge~(\ref{eq:18}), one gets:
\begin{equation}\label{eq:22}
\alpha_s^{INP}(q^2)=\alpha_s^{INP}(z) = {z_c \over z},  \ \ \textrm{where} \ \ z =
{q^2 \over q_{eff}^2},  \ \ \textrm{and} \ \ z_c= {\Delta^2 \over
q_{eff}^2},
\end{equation}
so that the dimensionless effective potential~(\ref{eq:17}) becomes,
\begin{equation}\label{eq:23}
\Omega_{YM} (z_c) = {1 \over \pi^2} \int_0^1 \dd z \ z \left[ \ln
[1 + (3 z_c / z)] - {3\over 4 }{z_c \over z} \right].
\end{equation}
Performing an almost trivial integration in this integral, one obtains
\begin{equation}\label{eq:24}
\Omega_{YM} (z_c) = {1 \over 2 \pi^2} z^2_c \left[ {3 \over 2 z_c}
+ {1 \over z_c^2} \ln \left(1 + 3 z_c \right) - 9 \ln \left(1 + {1
\over 3 z_c} \right) \right].
\end{equation}
It is easy to see now that as a function of $z_c$, the effective
potential~(\ref{eq:24}) approaches zero from above as $\sim z_c$ at $z_c
\rightarrow 0$ limit. At infinity $z_c \rightarrow \infty$ it diverges
as $\sim - z_c$. At a fixed effective scale $q^2_{eff}$ and from
Eq.~(\ref{eq:22}) it follows that $z_c \rightarrow 0$ is a correct PT
regime, while $z_c \rightarrow \infty$ is not a physical regime, since
the mass gap $\Delta^2$ is either finite or zero (the PT limit), i.e.,
it cannot be infinitely large. In other words, at a fixed effective
scale one recovers the correct PT limit for the bag constant, i.e.,
the above-mentioned normalization condition is maintained for the bag
constant, as it should be.

\begin{figure}
\begin{center}
\includegraphics[width=12.0truecm,height=10.0truecm]{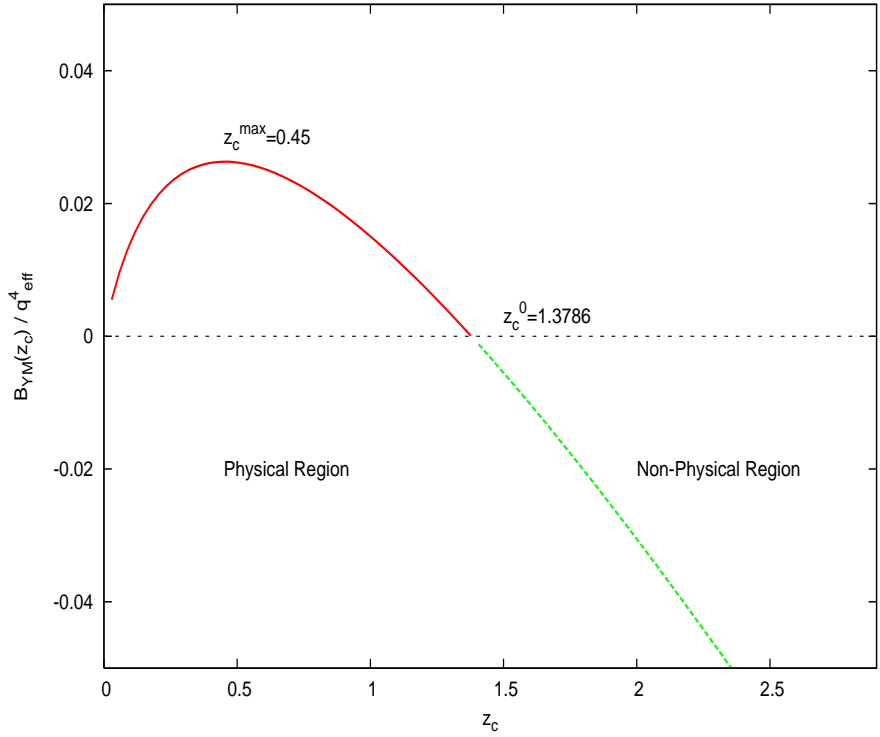}
\caption{The $B_{YM} / q^4_{eff}$ effective potential vs. $z_c$.
The non-physical region is $z_c \geq z_c^0$, since $B_{YM}$ should
be always positive. At $z_c=0$ the effective potential is also
zero.} \label{fig:3}
\end{center}
\end{figure}

\begin{figure}
\begin{center}
\includegraphics{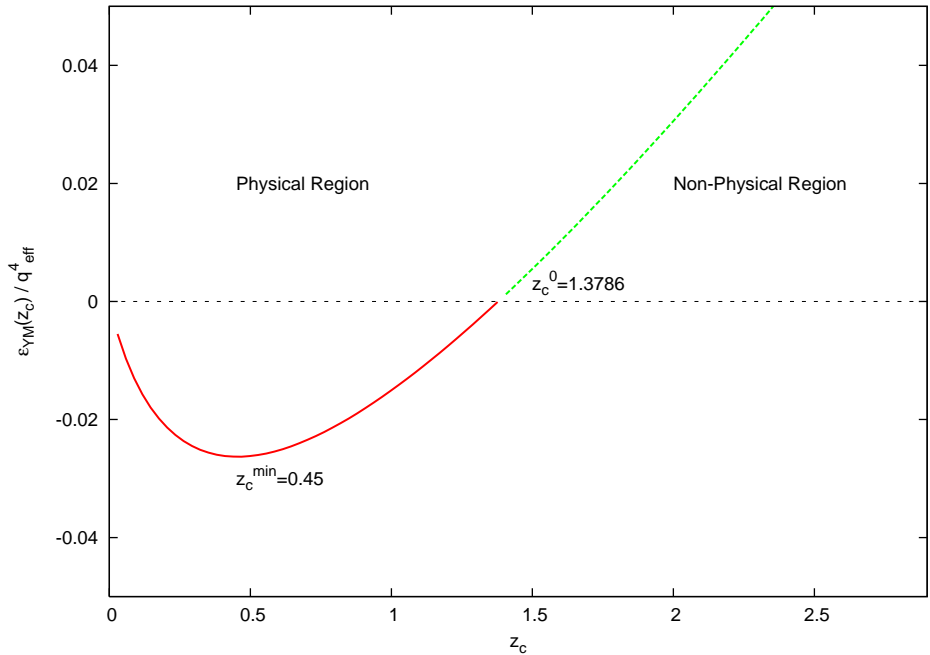}
\caption{The $\epsilon_{YM} / q^4_{eff}$ effective potential vs.
$z_c$. The non-physical region is $z_c \geq z_c^0$, since
$\epsilon_{YM}$ should be always negative. At $z_c =0$ the
effective potential is also zero.} \label{fig:4}
\end{center}
\end{figure}

The nontrivial second zero of the effective potential~(\ref{eq:24})
follows obviously from the condition,
\begin{equation}\label{eq:25}
3 z_c  + 2 \ln (1 + 3 z_c ) - 18 z_c^2 \ln \left(1 + (1 / 3 z_c)
\right) =0,
\end{equation}
which numerical solution is
\begin{equation}\label{eq:26}
z_c^0 = 1.3786.
\end{equation}
Evidently, through the relation~(\ref{eq:22}) this value determines a
possible upper bound for $\Delta^2$ and lower bound for $q_{eff}^2$,
since $B_{YM}/ \epsilon_{YM}$ is always positive/negative (see
Figs.~\ref{fig:3} and \ref{fig:4}).

At $z_c=0$, i.e., $\Delta^2=0$ the effective potential~(\ref{eq:24})
vanishes identically, as it should be. From the above one can conclude
that this effective potential as a function of $z_c$ has a maximum at
some finite point, see Fig.~\ref{fig:3}. In the way how it has been
introduced $z_c$ plays the role of the constant of integration of the
effective potential though being formally a parameter of the theory.
In general, by taking the first derivative of the effective potential with
respect to the constant of integration one recovers the corresponding
equations of motion~\cite{13,14,15}. Requiring thus $\partial
\Omega_{YM} (z_c) / \partial z_c = 0$, one obtains:
\begin{equation}\label{eq:27}
z_c^{-1} = 4 \, \ln [1 + (1 / 3z_c)],
\end{equation}
which makes it possible to fix the constant of integration of the
corresponding equation of motion at maximum. Its numerical solution is
\begin{equation}\label{eq:28}
z_c^{max} = 0.4564 < 1,
\end{equation}
so at maximum the ratio $\Delta^2 / q^2_{eff}$ is always less than one.
At this point the numerical value of the effective
potential~(\ref{eq:24}) is
\begin{equation}\label{eq:29}
\Omega_{YM} (z_c^{max}) = { 1 \over 2 \pi^2} \left[ {3 \over 4}
(z_c^{max}) -  \ln \left(1 + 3 z_c^{max} \right) \right] = 0.0263.
\end{equation}
The bag constant defined in Eq.~(\ref{eq:16}), and hence the
corresponding INP VED~(\ref{eq:13}), as a function of $q^4_{eff}$
or, equivalently, of the mass gap $\Delta^4$ thus becomes,
\begin{equation}\label{eq:30}
B_{YM} = - \epsilon_{YM} = 0.0263 \, q_{eff}^4 = 0.1273 \times
\Delta^4,
\end{equation}
where the relation
\begin{equation}\label{eq:31}
q_{eff}^2 = (z_c^{max})^{-1} \Delta^2 = 2.2 \ \Delta^2
\end{equation}
has been already used. It is worth noting that a maximum for the
bag constant corresponds to a minimum for the INP YM VED
$\epsilon_{YM}$ (the so-called "stationary" state, see
Fig.~\ref{fig:4}).

So, we have explicitly demonstrated that in the considered case
the bag constant~(\ref{eq:30}) is finite, positive, and it has no
imaginary part, indeed. It depends only on the mass gap responsible
for the INP dynamics in the QCD ground state or, equivalently, on
the effective scale squared separating the NP region from the PT one.

\subsection{Scale-setting schemes and numerical results}
\label{sec:numres}

In order to complete the numerical calculation of the above defined
bag constant all we need now is the value for the effective scale
$q^2_{eff}$, which separates the NP region from the PT one.
Similarly, the value for a scale at which the NP effects become
important, that is the mass gap $\Delta^2$, also allows one to
achieve the same goal. If the PT regime for gluons (as well as for
quarks) starts conventionally from $1 \ \GeV$, then this number is
a natural choice for the effective scale. It makes it also possible
to directly compare our values with the values of many
phenomenological parameters calculated just at this scale (see below).
We consider this value as well justified and realistic upper limit
for the effective scale defined above. Thus, using further the
relation~(\ref{eq:31}), one gets
\begin{equation}\label{eq:32}
q_{eff}^2 = 1 \ \GeV^2, \ \ \ \textrm{and} \ \ \ \Delta^2 = 0.4564 \ \GeV^2.
\end{equation}
Similarly, the numerical value of the mass gap $\Delta^2$ has been
obtained from the experimental value for the pion decay constant,
$F_{\pi} =93.3 \ \MeV$, by implementing a physically well-motivated
scale-setting scheme~\cite{44,45}. In fact, we approximate the pion
decay constant in the chiral limit $F^0_{\pi}$ by its experimental
value, since the difference between them can be a few MeV only. This
is due to smallness of the corresponding light quark current masses.
The pion decay constant is a good experimental number, since it is
directly measured quantity, contrary to, for example the quark
condensate or the dynamically generated quark mass. For the mass gap
we have obtained the following numerical result $\Delta = 0.5784 \ \GeV$,
so similarly to the relations~(\ref{eq:32}), one yields

\begin{equation}\label{eq:33}
\Delta^2 = 0.3345 \ \GeV^2, \ \ \ \textrm{and} \ \ \ q_{eff}^2 = 0.733 \ \GeV^2.
\end{equation}
In what follows we will consider this value as a realistic lower
limit for the effective scale. One has to conclude that we have
obtained rather close numerical results for the effective scale
and the mass gap, by implementing rather different scale-setting
schemes. It is worth emphasizing that the effective scale~(\ref{eq:33})
covers quite well not only the deep IR region but the substantial
part of the intermediate one as well.

For the above-mentioned possible upper bounds for $\Delta^2$ and
lower bounds for $q^2_{eff}$ our numerical results are for the
scale-setting scheme~(\ref{eq:32}):
\begin{equation}\label{eq:34}
\Delta^2 \leq 1.379 \ \GeV^2, \ \ \ \textrm{and} \ \ \ q^2_{eff} \geq 0.330 \ \GeV^2,
\end{equation}
then similarly, based on the scale-setting scheme~(\ref{eq:33}):
\begin{equation}\label{eq:35}
\Delta^2 \leq 1.01 \ \GeV^2, \ \ \ \textrm{and} \ \ \ q^2_{eff} \geq 0.242 \ \GeV^2.
\end{equation}
Evidently, their calculated values in each scale-setting scheme
satisfy the corresponding bounds.

For the bag constant (and hence for the INP YM VED) from
Eq.~(\ref{eq:30}), one obtains
\begin{equation}\label{eq:36}
B_{YM} = - \epsilon_{YM} = (0.0142-0.0263) \ \GeV^4,
\end{equation}
where the first and second numbers in brackets correspond to the
numerical values given in Eqs.~(\ref{eq:32}) and~(\ref{eq:33}),
respectively.

In conclusion, let us note that in the pure YM theory there is no way
to calculate the mass gap independently of the well-motivated
scale-setting scheme, that's the effective scale in this case, i.e.,
relations~(\ref{eq:32}). The scale-setting scheme~(\ref{eq:33}) is
based on the numerical value of the pion decay constant in the
chiral limit. So this scheme is legitimated to use here as well,
since the chiral quark condensates do not
contribute to the VED in this limit, as it follows from the trace anomaly relation
(see next section). For further discussion on the numerical value
of $B_{YM}$ in different units see appendix B.

\section{The trace anomaly relation}
\label{sec:tarce}

The TNP VED (and hence the bag constant) is important by itself as
the main dynamical characteristic of the QCD ground state.
Furthermore it assists in calculating such an important
phenomenological parameter as the gluon condensate, introduced in
the QCD sum rules approach to the physics of resonances~\cite{11}.
The famous trace anomaly relation~\cite{16,17,18,19} in the general
case of non-zero current quark masses $m_f^0$ is
\begin{equation}\label{eq:37}
\Theta_{\mu\mu} = {\beta(\alpha_s) \over 4 \alpha_s} G^a_{\mu\nu}
G^a_{\mu\nu} + \sum_f m_f^0 \overline q_f q_f,
\end{equation}
where $\Theta_{\mu\mu}$ is the trace of the energy-momentum tensor
and $G^a_{\mu\nu}$ being the gluon field strength tensor, while for
the ratio $\beta(\alpha_s) / \alpha_s$ see discussion below. The
trace anomaly relation which includes the anomalous dimension for
the quark mass has been derived in Ref.~\cite{19}, however, in our
case of the pure gluon fields we can use the standard form of the
trace anomaly relation~(\ref{eq:37}). Sandwiching it between vacuum
states and taking into account the obvious relation
$\langle{0} | \Theta_{\mu\mu} | {0}\rangle = 4 \epsilon_t$, one
obtains
\begin{equation}\label{eq:38}
4 \epsilon_t =  \langle{0} | {\beta (\alpha_s)  \over 4 \alpha_s}
G^a_{\mu\nu} G^a_{\mu\nu} | {0}\rangle +  \sum_f m^0_f \langle{0}
| \overline q_f q_f | {0}\rangle.
\end{equation}
Here $\epsilon_t$ is the sum of all possible independent NP
contributions to the VED (the total VED) and $\langle{0} |
\overline q_f q_f | {0}\rangle$ is the chiral quark condensate.
From this equation in the case of the pure YM fields (i.e., when
the number of quark fields is zero $N_f=0$), one can get
\begin{equation}\label{eq:39}
\langle{0} | {\beta(\alpha_s) \over 4 \alpha_s} G^a_{\mu\nu}
G^a_{\mu\nu} | {0}\rangle = 4 \, \epsilon_{YM},
\end{equation}
where, evidently we saturate the total VED, $\epsilon_t$ by the TNP
YM VED, $\epsilon _{YM}$ defined in Eq.~(\ref{eq:13}), i.e., putting
$\epsilon_t = \epsilon_{YM} + ...$. Let us note that the same result,
i.e., Eq.~(\ref{eq:39}), will be obtained in the chiral limit for
light quarks $m_f^0=0$, for $f=1,2,3$ as well.

If confinement happens then the $\beta$ function is always in the
domain of attraction (i.e., always negative) without IR stable fixed
point~\cite{1}. Therefore, it is convenient to introduce the general
definition of the gluon condensate not using the weak coupling limit
solution to the $\beta$ function as follows:
\begin{equation}\label{eq:40}
\langle G^2 \rangle \equiv  - \langle{0} | {\beta(\alpha_s) \over
4 \alpha_s} G^a_{\mu\nu} G^a_{\mu\nu} | {0}\rangle =  - 4 \,
\epsilon_{YM} = 4 \, B_{YM}.
\end{equation}
Thus, the above defined general gluon condensate will be always
positive, as it should be. The importance of this relation is that
it gives the value of the gluon condensate as a function of the bag
constant whatever solution of the $\beta$ function in terms of
$\alpha_s$ is. However, let us remind that there is a correlation
between the two sides of this equation. The bag constant, correctly
defined in Eq.~(\ref{eq:13}), depends, in general, on the TNP
effective charge $\alpha_s^{TNP}(q^2)$. On the other hand, the
renormalization group equation
\begin{equation}\label{eq:41}
q^2 \ {\dd \alpha_s(q^2) \over \dd q^2} = \beta(\alpha_s(q^2))
\end{equation}
for the $\beta$ function gives it in terms of the corresponding
effective charge. This makes it possible to determine the ratio
$(\beta(\alpha_s) / \alpha_s) \equiv (\beta(\alpha_s(q^2)) /
\alpha_s(q^2))$, which appears in the left-hand-side of
Eq.~(\ref{eq:40}). Of course, this equation should be solved for the
chosen TNP effective charge (see subsection 7.1).

Concluding, let us only note that the quantum part of the total TNP
VED at log-loop level is:
\begin{equation}\label{eq:42}
\epsilon_t = \epsilon_{YM} + N_f \, \epsilon_q,
\end{equation}
where $\epsilon_q$ is the TNP quark skeleton loop contribution, see
the corresponding skeleton loop diagram in Fig.~\ref{fig:1}. It is an
order of magnitude less than $\epsilon_{YM}$ because of much less
quark degrees of freedom in the vacuum, and it is positive because of
overall minus due to the quark loop. Evidently, in terms of the YM
bag constant, one obtains
\begin{equation}\label{eq:43}
\epsilon_t = - B_{YM}\left[1 - \nu N_f \right],
\end{equation}
where we introduce $\epsilon_q = \nu B_{YM}$ and $\nu \ll 1$. So the
replacement of the total bag constant by its YM counterpart only is
a rather good approximation from the numerical point of view. In this
connection, let us remind that in the large $N_c$-limit the pure
gluon contribution scales as $N_c^2$, while the quark contribution
scales only as $N_c$~\cite{46}. However, in order to correctly
calculate the bag constant in full QCD the quark part of the TNP
VED $\epsilon_q$ is also important. Let us note that it is non-zero even in the chiral limit.
This part will be investigated and calculated in our subsequent paper.

\subsection{Comparison with phenomenology}
\label{sec:comp}

Let us show explicitly now that our numerical values for the bag
constant calculated in~(\ref{eq:36}) are in rather good agreement with
the phenomenological values of the gluon condensate. Above we have
already developed a general formalism which allows one to express the
gluon condensate as a function of the bag constant. So substituting the
numerical value of the bag constant into the Eq. (40), one obtains:
\begin{equation}\label{eq:44}
\langle G^2 \rangle \equiv  - \langle{0} | {\beta(\alpha_s) \over
4 \alpha_s} G^a_{\mu\nu} G^a_{\mu\nu} | {0}\rangle = 4 \, B_{YM} =
(0.0568-0.1052) \ \GeV^4.
\end{equation}
On the other hand, the renormalization group equation for the $\beta$
function~(\ref{eq:41}) after substitution of our solution for the INP
effective charge~(\ref{eq:18}) yields:
\begin{equation}\label{45}
\beta(\alpha_s(q^2)) = - \alpha_s(q^2),
\end{equation}
as it is required for the confining theory where the $\beta$ function
should be always in the domain of attraction, i.e., negative (see in
Ref.~\cite{1}). The corresponding ratio as it appears in the
left-hand-side of Eq.~(\ref{eq:44}) is
\begin{equation}\label{eq:46}
{\beta(\alpha_s) \over \alpha_s} \equiv {\beta(\alpha_s(q^2))
\over \alpha_s(q^2)} =-1.
\end{equation}
Substituting further this solution into the Eq.~(\ref{eq:44}), it becomes
\begin{equation}\label{eq:47}
\langle{0} | {1 \over 4 } G^a_{\mu\nu} G^a_{\mu\nu} | {0}\rangle =
4 \, B_{YM} = (0.0568-0.1052) \ \GeV^4,
\end{equation}
which means that both sides of this relation between the Bag constant
and the gluon condensate have been calculated by using the same
expression for the INP effective charge, and hence for the corresponding
$\beta$ function. So from the numerical point of view the Bag constant
and the gluon condensate are in a self-consistent dependence from each
other, making thus the latter one free of all the types of the PT
contributions. Our expression for the gluon condensate~(\ref{eq:47})
allows one to recalculate any gluon condensate at any scale and any
ratio, $\beta(\alpha_s) / \alpha_s$. To the gluon condensate a physical
meaning can be indeed assigned as the global (average) vacuum
characteristic which measures a density of the TNP gluon fields
configurations in the QCD vacuum.

However, it cannot be directly compared with the phenomenological
values for the standard gluon condensate estimated within different
approaches~\cite{20}. The problem is that it is necessary to remember
that any value at the scale as in Eq.~(\ref{eq:33}) (lower bound in
the right-hand-side of Eq.~(\ref{eq:47})) is to be recalculated at the
$1 \ \GeV$ scale. Moreover, both values explicitly shown in
Eq.~(\ref{eq:47}) should be recalculated at the same ratio, as
mentioned above.

In phenomenology the standard ratio of the gluon condensate and its
numerical value is:
\begin{equation}\label{eq:48}
G_2 = \langle {\alpha_s \over \pi} G^2 \rangle = \langle{0} |
{\alpha_s \over \pi} G^a_{\mu\nu} G^a_{\mu\nu} | {0}\rangle
\approx 0.012 \ \GeV^4 \ ,
\end{equation}
which can be changed within a factor of $\sim 2$~\cite{11} (let us
recall that this ratio comes from the weak coupling solution for
the $\beta$ function, see for example in Ref.~\cite{47}).

Thus in order to achieve the same ratio the both sides
of Eq.~(\ref{eq:47}) should be multiplied by
$4(\alpha_s / \pi)$. For the
numerical value of the strong fine structure constant we use
$\alpha_s = \alpha_s(m_Z) = 0.1187$ from the Particle Data
Group~\cite{48}. In addition, the lower bound should be multiplied
by the factor $ (1 / 0.733)^2 =1.86$, coming form the numerical
value by Eq.~(\ref{eq:33}). Then the recalculated gluon condensate
in~(\ref{eq:47}), which is denoted as $\bar G_2$, finally becomes (i.e., both numbers
in Eq.~(\ref{eq:47}) coincides, as it should be)
\begin{equation}\label{eq:49}
\bar G_2 \approx 0.016 \ \GeV^4.
\end{equation}
This numerical value for the gluon condensate should be compared
with the numerical value coming from the phenomenology, see
Eq.~(\ref{eq:48}) above. This shows that all our numerical results are
in good agreement with various phenomenological estimates~\cite{11,20}, taking into account that
the quark contributions are approximately an order of magnitude less than the pure YM one to the full bag constant
(see remarks in this section just before subsection 7.1). This confirms that our numerical values for the bag
constant and hence for the gluon condensate are rather realistic ones.

\section{Conclusions}
\label{sec:concl}

In summary, we have formulated a general method how to calculate
numerically the quantum part of the TNP YM VED
(the YM bag constant, apart from the sign, by definition)
in the covariant gauge QCD from first principles. For this purpose we have used the
effective potential approach for composite operators to leading order \cite{13}. It has an
advantage to be directly the VED (the pressure) in the absence of
external sources. The bag constant is defined as a special function
of the TNP effective charge integrated out over the NP region (soft
momentum region), see Eq.~(\ref{eq:13}). At this stage the bag
constant is colorless (color-singlet) and depends only on the
transversal ("physical") degrees of freedom of gauge bosons. It
is also free of the PT contributions by its construction. This has
been achieved due to the subtractions at the fundamental
level as given by Eg.~(\ref{eq:7}), as well as due to all other
subtractions explicitly shown in Eq.~(\ref{eq:12}), when the gluon
degrees of freedom were to be integrated out. Thus, our equations
(\ref{eq:16}) and (\ref{eq:17}) are general ones in order to
correctly calculate the bag constant as a function
of any properly defined TNP effective charge.

For the concrete calculation of the bag constant we replace the TNP effective
charge by its confining INP counterpart in Eq.~(\ref{eq:18}), since
it is exactly and uniquely separated from the PT effective charge.
The INP effective charge depends regularly on the mass gap, which
is responsible for the large-scale structure of the QCD ground
state~\cite{24,34}. The scale-setting schemes have been chosen by
the two different ways, leading, nevertheless, to a rather close
numerical results for the mass gap and hence for the effective scale.
The calculated bag constant in addition, is: finite, positive, and
it has no imaginary part (stable vacuum). It is also a manifestly
gauge-invariant quantity (i.e., does not explicitly depend on the
gauge-fixing parameter as it is required). The separation of "soft
versus hard" gluon momenta is also exact because of the
maximization/minimization procedure. It becomes possible since the
effective potential ~(\ref{eq:24}) as a function of the constant of
the integration $z_c$ and hence of the mass gap $\Delta^2$ has a local maximum, see Fig.~\ref{fig:3}.
This also makes it possible that in the above-mentioned scale-setting
schemes either the mass gap or the effective scale is only independent,
since the other one is to be determined via the relation~(\ref{eq:31}).
In the scale-setting scheme~(\ref{eq:32}) the effective scale is
independent, while in the second scale-setting scheme~(\ref{eq:33})
the mass gap is independent. {\it It is worth emphasizing that the
bag constant in our approach is not simply the difference
between the PT and NP VEDs, which is finite, colorless and
manifestly gauge-invariant, etc. It is the energy density (apart from the sign) of the system
of stable configurations of the purely transversal quantum virtual
fields with the enhanced low-frequency components/large scale
amplitudes due to the NL interaction of massless gluon modes, and
which is being at "stationary state", i.e., being in the state with the minimum of energy,
see Fig.~\ref{fig:4}}.

In order to compare our numerical results with phenomenology we
develop a general formalism which makes it possible to relate the
bag constant to the gluon condensate in a unique and self-consistent
way. In other words, the gluon condensate is defined and calculated at the same effective charge,
which has been chosen for the calculation of the bag constant.
For this purpose we use the trace anomaly relation without
applying to the weak coupling solution for the corresponding $\beta$
function. In its turn, it is a solution of the corresponding renormalization group equation
for the effective charge Eq.~(41). Our numerical results turned out to be in good agreement
with phenomenological values of the gluon condensate calculated and
estimated within different approaches and methods~\cite{11,20}.

It is instructive to briefly summarize our theoretical and numerical
results for the bag constant in general and our specific ways:

\vfill

\eject

\begin{description}
\item[General properties of the bag constant determined by Eqs.~(16)-(17) are:]
\end{description}
\begin{itemize}
\item colorless (color-singlet);
\item electrically neutral;
\item transversal, i.e., depending only on "physical" degrees of
freedom of gauge bosons;
\item free of the PT contributions ('contaminations').
\end{itemize}

\begin{description}
\item[Results, depending on the confining effective charge Eq.~(18) are:]
\end{description}
\begin{itemize}
\item the explicit gauge invariance;
\item uniqueness, i.e., it is free of all the types of the PT
contributions now;
\item finiteness;
\item positiveness;
\item no imaginary part (stable vacuum);
\item existence of the stationary state for the corresponding YM
energy density (negative pressure, see Fig. 4);
\item the final dependence on the mass gap only;
\item a good numerical agreement with phenomenology.
\end{itemize}

The above remarkable features all together are unique. Apparently, it
is due to the confining expression~(\ref{eq:18}) and the correct
determination of the bag constant itself in this investigation. It has
been made in accordance with its modern definition as the difference
between the PT and the NP VEDs~\cite{9,10,11,12}.

Our method can be generalized on the multi-loop skeleton contributions
to the effective potential approach for composite operators, as well
as to take into account the quark degrees of freedom, as plotted in
Fig.~\ref{fig:1}. These terms, however, will produce numerical
contributions an order of magnitude less, at least, in comparison with
the leading log-loop level gluon term given by Eq.~(\ref{eq:1}). What
is necessary indeed, is to be able to extract the finite part of the TNP VED in a
self-consistent and manifestly gauge-invariant ways. This is provided
by our method which thus can be applied to any QCD vacuum quantum and
classical models at any gauge (covariant or non-covariant). It may
serve as a test of them, providing an exact criterion for the
separation "stable versus unstable" vacua. Using our method we have
already shown that the vacuum of classical dual Abelian Higgs model
with string and without string contributions is unstable against
quantum corrections~\cite{49,50}.

It would be also interesting to apply our general equations~(\ref{eq:16})
and~(\ref{eq:17}) in order to calculate the bag constant within the
recently obtained analytical results for the gluon propagator in
Refs. \cite{51,52,53,54,55,56,57,58}. The general formalism developed in our paper is aimed first of all at the analytical calculations of the bag constant (or the vacuum energy density) in any model gluon propagator
in continuous QCD. However, as mentioned above the chosen ansatz (18) can be considered as useful parametrization of the corresponding lattice results. In this way our formalism can be extended to the lattice calculations as well. Choosing an appropriate parametrization of any lattice result for the gluon propagator
(there is a lot of recent lattice data \cite{59,60,61,62} and references therein), one then can substitute it into our analytical expressions (16)-(17). Such a combination of the lattice and analytical calculations can be rather effective indeed, in order to understand what is the physics behind the lattice numbers and curves. On the other hand, all the analytical expressions and calculations will be put on solid numerical ground provided by the lattice simulations. So there is no doubt that the analytical and lattice calculations should not exclude each other, but contrary they should complement each other. All these possible developments are, of course, beyond the scope of the present investigation, and they have to be done elsewhere.


\ack{Support in part by HAS-JINR and Hungarian OTKA-TO47050, NK62044
and IN71374 (P. L\'{e}vai), PD73596 (GGB) is to be acknowledged.
We are grateful to J. Nyiri for constant support and help. We
would like also to thank M. Vas\'uth for help in the numerical
calculations and \'A. Luk\'acs for useful discussion. This paper was partially supported by the J\'anos
Bolyai Researc Scholarship of the HAS.}

\appendix

\section{The general role of ghosts}
\label{AppendixA}

Let us begin with recalling that due to the above-mentioned
normalization condition in the initial Eq. (1), its elaborated
counterpart in Eq. (6) depends only on the transversal
("physical") component of the full gluon propagator. So there is
no need for ghosts to cancel its longitudinal (unphysical)
component, indeed. However, it is instructive to discuss the role
of ghosts in general, and to clearly show that their explicit
introduction leads to the same result for the bag constant, in
particular.

Following Ref.~\cite{13}, the effective potential at the same
log-loop order for the ghost degrees of freedom analytically is:
\begin{equation}\label{eq:A.1}
V(G) =  - i  \int {\dd ^4k \over {(2\pi)^4}}
 Tr\{ \ln (G_0^{-1}G) - (G_0^{-1}G) + 1 \},
\end{equation}
where $G \equiv G(k)= i / k^2(1 + b(k^2))$ is the full ghost
propagator, where $b(k^2)$ is the ghost self-energy, while $G_0
\equiv G_0(k) = i / k^2$ is its free PT counterpart. Trace over
color group indices is assumed. Evidently, the effective potential
is normalized to $V(G_0) = 0$ in the same way as the gluon part
in Eq.~(\ref{eq:1}). Substituting these expressions into the ghost
term~(\ref{eq:A.1}) and again doing some algebra in four-dimensional
Euclidean space, one formally obtains that
$V(G) = \epsilon_{gh} = \int \dd k^2 f(b(k^2))$. This, in general,
divergent constant contribution should be of course, regularized
in order to assign to it a mathematical meaning. So the explicit
functional dependence of the ghost propagator/self-energy on
its argument is not important, since within the effective
potential approach to calculate the VED it is always only
constant. We have to sum up all the contributions for the pure
YM fields at the same skeleton log-loop order. The
relation given by Eq.~(\ref{eq:10}) then should look like as:
\begin{eqnarray}\label{eq:A.2}
\epsilon_g + \epsilon_{gh} = &-& {1 \over \pi^2} \int_0^{q^2_{eff}}
\dd q^2 \ q^2 \left[ \ln [1 + 3 d^{TNP}(q^2)] - {3 \over
4}d^{TNP}(q^2) \right] + \nonumber \\
&+& \epsilon_{PT} + \epsilon'_{PT} + \epsilon_{gh}.
\end{eqnarray}
It is worth emphasizing that, the right-hand-side of this relation
may still suffer from unphysical singularities by the integral in
Eq~(\ref{eq:9}), defining $\epsilon_{PT}$. The problem is that the
PT effective charge, $d^{PT}(q^2)$, which is responsible for
AF in QCD at large $q^2$ (see, for example our
paper~\cite{34}), may have, in general, unphysical singularities
below the scale $\Lambda^2_{QCD}$, since in the integral~(\ref{eq:9})
the integration is from zero to infinity. In addition, as mentioned
above the integral~(\ref{eq:11}), defining $\epsilon'_{PT}$, may
be still divergent. Thus the left-hand-side of the
relation~(\ref{eq:A.2}) is formal one, indeed. It suffers from various
types of unphysical singularities which may appear in its
right-hand-side. In order to get a physically meaningful expression,
one has to remove the two integrals~(\ref{eq:9}) and~(\ref{eq:11})
from Eq.~(\ref{eq:6}). This is to be done with the help of a ghost
term by imposing the following condition of cancelation of unwanted
terms in the most general form:
$\epsilon_{PT} + \epsilon'_{PT} + \epsilon_{gh}=0$. This condition
can be always fulfilled, since it is a relation between three
different (unknown in general) regularized constants. Then the relation~(\ref{eq:A.2}) thus becomes:

\begin{eqnarray}\label{eq:A.3}
\epsilon_{YM} &=& \epsilon_g - \epsilon_{PT} - \epsilon'_{PT} \nonumber \\
&=&
- {1 \over \pi^2} \int_0^{q^2_{eff}} \dd q^2 \ q^2 \left[ \ln [1 + 3
d^{TNP}(q^2)] - {3 \over 4}d^{TNP}(q^2) \right],
\end{eqnarray}
in complete agreement with the relation~(\ref{eq:12}), and hence
with the definition of the bag constant~(\ref{eq:13}), as it
should be. So the TNP gluon contribution to the VED has been
determined by subtracting unwanted terms by means of the ghost
contribution. Evidently, the subtracted terms are of no importance,
while a ghost term plays no explicit role for further consideration.

In QCD the general role of ghost degrees of freedom is to cancel
all the unphysical degrees of freedom of gauge bosons~\cite{1,63},
maintaining thus unitarity of the $S$-matrix. This is the main
reason why they should be taken into account together with gluons
always. This means that nothing should {\it explicitly} depend on them
after the above-mentioned cancelation is performed. One of the
main purposes of their introduction is to exclude the longitudinal
(unphysical) component of the gluon propagator in every order of
the PT, thus going beyond it and thus being a general one, indeed.
If there is no need to cancel the longitudinal component of gauge
boson propagators, then they should be used to eliminate the
unphysical singularities of gauge bosons below the QCD asymptotic
scale (as it was described above), or some other ones which may be
inevitably present in any solution/ansatz for the full gluon
propagator. If one knows the ghost propagator exactly, then the
above-mentioned cancelation of unphysical singularities of gauge
bosons should proceed automatically, as usual in the PT calculus
(if, of course, all calculations are correct). For such an exact
cancelation of the longitudinal part of the gluon propagator by
the free PT ghost propagator in lower order of the PT see, for
example Ref.~\cite{63}. But if it is not known exactly or known
approximately (depending on the truncation/approximation scheme),
as usual in the NP calculus then nevertheless, one has to impose
the corresponding condition of cancelation in order to fulfill
their general role. This just has been done above. Thus our
subtraction scheme is in agreement with the general interpretation
of ghosts to cancel all the unphysical degrees of freedom of gauge
bosons~\cite{1,63}. So by themselves the ghosts cannot change the truly NP dynamics of QCD, associated
with the transversal component of the full gluon propagator in Eq.~(2) and described by its Lorentz
structure or, equivalently, by its effective charge (see Ref. \cite{25} as well).

Whatever solution(s) for the full gluon propagator
obtained by lattice QCD \cite{59,60,61,62} (and references therein) and by the analytical
approach based on the corresponding SD system of equations \cite{51,52,53,54,55,56,57,58}
(and references therein) might be (smooth, singular, massive, etc.), it, however, should
not undermine the above-mentioned general job of ghosts. It is worth emphasizing that
by no coincidence in all the papers cited above the transversal
Landau gauge has been chosen by hand from the very beginning. So
there is {\it no and cannot be the explicit} dependence on the
ghost degrees of freedom in any expressions for the physical
quantities, in general, and in the expression for the bag constant,
in particular. In this connection, let us remind that the confining
effective charge~(\ref{eq:18}) is the effective charge of the
relevant gluon propagator, which becomes the purely transversal in
a gauge invariant way, by construction~\cite{33,34}.

Nevertheless, this does not mean that we need no ghosts at all.
First of all, we need them in the higher orders of the two-particle irreducible vacuum graphs in the skeleton loop expansion of the effective potential \cite{13}, since for them the simple normalization of the free PT vacuum to zero does not work. So the cancelation of unphysical gluon modes should proceed with the help of the ghost degrees of freedom, as it was described in this appendix above. It is necessary to understand that the transversality of the gluon propagator in the Landau gauge in order to correctly treat the PT QCD Green's functions without ghosts is not enough to insure unitarity of the $S$-matrix in QCD. The whole machinery of all the ST identities and the corresponding SD equations is still needed in order to insure the unitarity cancelations even in the Landau gauge.

For example, the quark ST identity, contains the so-called ghost-quark scattering kernel explicitly~\cite{1}. This kernel still makes an important contribution to the identity even if the gluon propagator is
transversal \cite{64,65}. Omitting ghosts at all in this identity, one will lose an important piece of information on the quark degrees of freedom themselves. As a result, any solution of the quark SD equation will suffer from unphysical singularities in the complex momentum plane. The problem is that via the quark-gluon vertex this equation will crucially depend on the term which comes from the identity even if the gluon propagator is transversal. The completely NP analysis of this identity on the basis of the double pole structure of the full gluon  propagator in the IR,
Eq.~(18), has been made in our earlier papers \cite{66,67}. We have derived the corresponding expression for the quark-gluon vertex following Ref. \cite{68} only in more sophisticated way (see Ref. \cite{32} as well).
We will take this result into account when we will directly calculate the confining quark contribution to the bag constant, as mentioned in section 7 just before subsection 7.1.


\section{Numerical values for $B_{YM}$ in different units }
\label{AppendixB}

In order to show explicitly what magnitude of numbers we are dealing
with, let us present our numerical value for the bag constant given
by Eq.~(\ref{eq:36}) in different units, namely:
\begin{eqnarray}\label{eq:B.1}
B_{YM} = - \epsilon_{YM} &=& (0.0142-0.0263) \ \GeV ^4 \nonumber\\
&=& (1.84-3.4) \ \GeV / \fm ^3 \nonumber\\
&=& (1.84-3.4) \times 10^{39} \ \GeV / \cm ^3.
\end{eqnarray}
This is a huge amount of energy stored in one cm$^3$ of the QCD
vacuum even in "God-given" units $\hslash =c=1$. Using the number
of different conversion factors (see, for example Ref.~\cite{63}
or the particle data group~\cite{48}) the bag constant can be
expressed in different systems of units (SI, CGS, etc.).

Taking further into account that
\begin{equation}\label{eq:B.2}
1 \ \GeV = 1.6 \times 10^{10} \textrm{J} = 4.45 \times 10^{-23} \
\textrm{GWh},
\end{equation}
from Eq.~(\ref{eq:B.1}) one finally gets ($1$ W $= 10^{-3}$ kW $= 10^{-6}$
MW $= 10^{-9}$ GW)
\begin{equation}\label{eq:B.3}
B_{YM} = (8.2 - 15 ) \times 10^{16} \ \textrm{GWh}/
\cm ^3
\end{equation}
or, equivalently,
\begin{equation}\label{eq:B.4}
E_{YM} = B_{YM} \ \cm ^3 = (8.2 - 15 ) \times
10^{16} \ \textrm{GWh} \sim 10^{17} \ \textrm{GWh}
\end{equation}
in familiar units of watt-hour (Wh). Let us note that if one puts
the effective scale squared as small as realistically possible
$q^2_{eff} = 0.242 \ \GeV^2$ (see Eq.~(\ref{eq:35})), then the
previous number will be only slightly changed, namely
$E_{YM} = B_{YM} \ \cm ^3 = (4.8 - 8.7 ) \times 10^{15} \ \textrm{GWh}$.
So both numbers still indicate a huge amount of the bag energy
$E_{YM}$ stored in one cm$^3$ of the QCD vacuum.

It is especially interesting to compare these numbers with the total production of primary
energy of the 25 EU countries in year 2004 which was \cite{69} (see also Ref. \cite{70})

\begin{equation}\label{eq:B.5}
E_t \sim 10.2 \ \textrm{PWh} = 10.2 \times 10^6 \ \textrm{GWh} \sim 10^7 \ \textrm{GWh},
\end{equation}
where $1 \ \textrm{PWh} = 1 \ \textrm{Petawatt-hour}$. Approximately $1/3$ of this energy was produced
by nuclear power plants \cite{69,70}. The huge difference between the numbers in Eqs.~(\ref{eq:B.4})
and~(\ref{eq:B.5}) is very impressive and leads to some interesting
still speculative but already possible discussion in appendix D below and in our preliminary work \cite{71}.

\section{Contribution of $B_{YM}$ to the dark energy problem}
\label{AppendixC}

Apparently, our bag constant~(\ref{eq:B.1}) may also contribute to the so-called dark
energy density~\cite{72}. At least, from the qualitative point of view
it satisfies almost all the criteria necessary for the dark
energy/matter candidate (see here section 8 and discussions in Refs.~\cite{72,73}).
From the quantitative numerical point of view it is also much better
than the estimate from the Higgs field's contribution to the VED,
which is about \cite{74,75}

\begin{equation}
\varrho_H \sim 10^8 \ \GeV^4.
\end{equation}
In this notation our value (B.1) is about

\begin{equation}
\varrho_{our} \sim 10^{-2} \ \GeV^4.
\end{equation}
The observed VED is very small indeed, namely

\begin{equation}
\varrho_{vac} \sim 10^{-46} \ \GeV^4,
\end{equation}
see Refs. \cite{74,75,76}. So relatively
to the value inferred from the cosmological constant (i.e., the above-mentioned observed VED)

\begin{equation}
\varrho_H / \varrho_{vac} \sim 10^{54},
\end{equation}
while our is

\begin{equation}
\varrho_{our} / \varrho_{vac} \sim 10^{44},
\end{equation}
i.e, some $10$ orders of magnitude better, which is expected from the direct comparison of the
estimate (C.1) with our value (C.2).

Let us note that calculating at the Plank length scale~\cite{48}, we will obtain the same ratio,
as it should be. From Eq.~(B.1) it follows that

\begin{equation}
\varrho_{our} \sim 10^{39} \ \GeV / \cm ^3 = 10^{-60} \ \GeV / L_p ^3,
\end{equation}
where we used $\cm = 10^{33} \ L_p$ and $L_p$ denotes the above-mentioned Plank length \cite{48}.
In this units the observed VED is

\begin{equation}
\varrho_{vac} \sim 10^{-46} \ \GeV^4 \sim  10^{-5} \ \GeV / \cm ^3 = 10^{-104} \ \GeV / L_p ^3,
\end{equation}
so that the ratio between (C.6) and (C.7) becomes again (C.5), indeed.
Of course, the ratio (C.5) still remains very large, but it is much better than
the ratio (C.4), as emphasized above. Other possibility how QCD can be related to the dark energy
puzzle has been described in Ref. \cite{77} (and references therein).

Concluding, the vacuum for which the value (C.3) has been measured should not be mixed with the vacuum
of any quantum field gauge theory. For the former one its energy is always positive (i.e., above zero), so the vacuum is simply treated as empty space. The energy of the latter one is always negative (i.e., below zero), and it is full of any kind of quantum excitations, fluctuations, etc. However, the QCD bag constant is always positive, finite, gauge-invariant, etc. (if it has been correctly defined and calculated like in this work). That is the primary reason why we can compare our value (C.2) and the estimate (C.1) with (C.3).

\section{Energy from the QCD vacuum }
\label{AppendixD}

The Lamb shift and the Casimir effect are probably the two most
famous experimental evidences of zero-point energy fluctuations in
the vacuum of Quantum Electrodynamics (QED) \cite{78,79,80,81}. Both
effects are rather weak, since the QED vacuum is mainly PT by
origin, character and magnitude (the corresponding fine structure
constant is weak). However, even in this case attempts have been
already made to exploit the Casimir effect in order to "observe" the negative energy
and related affects \cite{80} and even to release
energy from the vacuum (see, for example Refs. \cite{82,83} and
references in the above-mentioned reviews \cite{78,79}). In Ref. \cite{84} by
investigating the thermodynamical properties of the quantum vacuum it has been
concluded that no energy can be extracted cyclically from the
vacuum (see, however Ref. \cite{78} and references therein). Let us also note that in QED
the photon propagator always remains PT even "dressed" \cite{25,34,85,86}. So formally we can define the bag constant in this theory as $B_{QED} = VED^0 - VED = - VED > 0$, since $VED^0 \equiv VED(D_0)=0$ in the effective potential approach to leading order \cite{13}. It would be interesting to perform such a calculation,
which will give one a correct finite value of the VED in QED, if, or course, the
above proposed definition of the QED bag constant makes sense. But it is beyond the scope
of the present investigation, and should be done elsewhere.

Since the QCD fine structure constant is strong, the idea to
exploit the QCD vacuum in order to extract energy from it seems
to be more attractive. However, before discussing the ways how to
extract, it is necessary to discuss which minimum/maximum amount
of energy at all can be released in a single cycle. Who
thinks that it is too early to discuss such kind of topic (though
we do not think so) may entirely skip this appendix.

The bag constant calculated here is a manifestly gauge-invariant, real
and colorless (color-singlet) quantity, i.e., it can be considered as a physical
quantity. In fact, in this paper we have formulated a
renormalization program to make the bag constant or, equivalently,
the bag pressure finite and satisfying all other necessary
requirements (see section 8 above). The key elements of this program were the necessary
subtractions at all levels. Moreover, one of its attractive
features, as emphasized above, is that it is the energy density of
the purely transversal virtual gluon field configurations which
are not only stable (no imaginary part),
but are being in the stationary state as well, i.e., in the
state with the minimum of energy (see Fig. 4). That is
why it makes sense to discuss the "releasing" of the bag constant
from the vacuum, more precisely the bag energy (B.4).

From the quantum statistical mechanics point of view, the energy
is nothing but the pressure multiplied by the volume $V$ in the
infinite-volume limit \cite{87}.  So the vacuum energy $E_{vac}$
in terms of the bag constant is and in $\GeV$ units it diverges as
follows:

\begin{equation}
E_{vac} = - B_{YM} \ V = - E_{YM}  {V \over \cm^3 } \sim -
\lambda^3, \quad \lambda \rightarrow \infty,
\end{equation}
since $V / \cm^3 \sim \lambda^3$ always when the dimensionless UV
cutoff $\lambda$ goes to infinity. Evidently, in deriving Eq. (D.1)
we use the general relation $E_{YM} = B_{YM} \ \cm^3$, which is valid in any units for energy
(see appendix B above).

Let us imagine now that we can release the finite portion $E_{YM}$
(B.4) from the vacuum in $k$ different places (different "vacuum
energy releasing facilities" (VERF)). It can be done by $n_m$
times in each place, where $m=1,2,3...k$. Then the releasing
energy $E_r$ becomes

\begin{equation}
E_r = E_{YM} \sum_{m=1}^k n_m.
\end{equation}
The ideal case (which, however, will never be achieved) is when we
could extract a finite portion of the energy an infinite number of
times and in an infinite number of places. So the releasing energy
(D.2) might be divergent as follows:

\begin{equation}
E_r = E_{YM}  \times \lim_{(k, n_m) \rightarrow \infty}
\sum_{m=1}^k n_m \sim \lambda^2, \quad \lambda
\rightarrow \infty,
\end{equation}
since the sum over $m$ diverges quadratically in the $\lambda
\rightarrow \infty$ limit, and $k \sim \lambda, \ n_m \sim
\lambda$ in this case. The difference between the vacuum energy
(D.1) and the releasing energy (D.3) which is nothing but the
remaining in the vacuum energy $E_R$ becomes

\begin{equation}
E_R = E_{vac} - E_r =  E_{vac}[ 1 + O(1/ \lambda)], \quad \lambda
\rightarrow \infty,
\end{equation}
i.e., the QCD vacuum is an infinite and permanent reservoir of
energy. The situation is even "better" if one takes into account
the PT contributions to the vacuum energy (in this case the
convergence becomes of the order $O(1/ \lambda^2)$ in Eq. (D.4),
see our preliminary work in Ref. \cite{71}).

That's the vacuum energy is badly divergent is not a mathematical
problem. This reflects an universal reality. Vacuum is everywhere
and it always exists. Quite possible that our Universe in general
and our real word in particular is only its special type of
excitation due to the Big Bang. As underlined above, the vacuum is
an infinite and hence a permanent source of energy. The only problem
is how to release the finite portion -- the bag energy (B.4) and whether it will be profitable or not
by introducing some type of cyclic process. However, due to huge
difference between the two numbers (B.4) and (B.5) such a
cyclically profitable process may be realistic. {\it "Perpetuum
mobile" does not exist, but "perpetuum source" of energy does
exist, and it is the QCD ground state}.



\end{document}